\documentclass[fleqn,usenatbib]{mnras}

\usepackage{newtxtext,newtxmath}

\usepackage[T1]{fontenc}
\usepackage{ulem}

\DeclareRobustCommand{\VAN}[3]{#2}
\let\VANthebibliography\thebibliography
\def\thebibliography{\DeclareRobustCommand{\VAN}[3]{##3}\VANthebibliography}

\usepackage{graphicx}	
\usepackage{amsmath}	

\newcommand{\Porb}{$P_{\rm orb}$}

\newcommand{\Mdot}{$\dot{M}$}

\newcommand{\Lx}{$L_{\rm X}$}

\newcommand{\src}{[HP99]\,159}

\newcommand{\ergs} {erg s$^{-1}$}

\newcommand{\kms}{\mbox{$\mathrm{km~s^{-1}}$}}
\newcommand{\he}[1] {He\,{\sc #1}}

\newcommand{\kT}{${\rm k}T$}

\title[First SSS with an evolved He donor]{\src\ -- Properties of the first Supersoft X-ray Source with a Helium star donor}

\author[H. Szegedi et al.]{H\'el\`ene Szegedi$^{1}$\thanks{E-mail: SzegediH@ufs.ac.za},
Philip A. Charles$^{1,2,3}$,
David A. H. Buckley$^{1,4,5}$,
Pieter J. Meintjes$^{1}$,
Przemek Mr\'oz$^{6}$,
\newauthor and Andrzej Udalski$^{6}$
\\
$^{1}$Department of Physics, University of the Free State, 205 Nelson Mandela Drive, Bloemfontein, 9300, RSA\\
$^{2}$Department of Physics and Astronomy, University of Southampton, Southampton, Hampshire, SO17 1BJ, United Kingdom\\
$^{3}$Astrophysics, Department of Physics, University of Oxford, Keble Road, Oxford OX1 3RH, UK\\
$^{4}$South African Astronomical Observatory, PO Box 9, Observatory Road, Observatory 7935, South Africa\\
$^{5}$Department of Astronomy, University of Cape Town, Private Bag X3, Rondebosch 7701, South Africa\\
$^{6}$Astronomical Observatory, University of Warsaw, Al. Ujazdowskie 4, 00-478 Warszawa, Poland\\
}

\date{Accepted XXX. Received YYY; in original form ZZZ}

\pubyear{\the\year{}}

\begin{document}
\label{firstpage}
\pagerange{\pageref{firstpage}--\pageref{lastpage}}
\maketitle

\begin{abstract}
\src\ is remarkable as the first supersoft X-ray source (SSS) identified with an evolved helium star donor. With a likely orbital period of 1.164\,d or 2.327\,d, the origin of the SSS component is controversial, with the two current models being either steady He-burning on the white dwarf surface, or that it is a helium nova in the decaying phase. To help resolve this issue we present extensive new long-term spectroscopy (with SALT) and photometry (at SAAO and with OGLE) of \src\ which (a) supports 2.327\,d as the orbital period, and (b) finds only a small \he{ii} radial velocity modulation. The latter is surprising as it implies a very low inclination system, whereas our light curve modelling suggests $i{\sim}50^\circ$, and hence that the \he{ii} must be produced in outflowing material further above, or beyond, the disc. We find that the decaying nova model cannot fit our OGLE light curve and the observed SSS flux level. \src\ has been essentially constant as an SSS over several decades, implying a sustained high level of mass-transfer from its He star donor, making it the only confirmed single-degenerate scenario SN Ia progenitor. We have updated the known SSS binary parameters and find a clear $\sim$1.5\,mag difference in their $M_{\rm V}$ when compared to the $M_{\rm V} - \Sigma$ properties of LMXBs, likely due to the larger irradiated areas and more luminous donors. 
\end{abstract}




\section{Introduction}

In spite of being detected in X-ray surveys over the last  $\sim$30 years, \src\ was only recently recognised as a new LMC X-ray source by \cite{Greiner2023}.  Based on its very soft X-ray spectrum (blackbody \kT\ $\sim$40-45\,eV, with a radius of $\sim$3700\,km) and relatively high luminosity (\Lx\ $\sim 7 \times 10^{36}$ \ergs), they suggested that it was a new member of the SSS (``supersoft'' X-ray sources) class \citep{Kahabka2006}, where the emission arises from ``steady'' thermonuclear burning of material on the surface of a white dwarf (WD).  What made \src\ remarkable,  though, was that its optical counterpart showed no hydrogen lines in its spectrum, only helium, leading \cite{Greiner2023} (hereafter, G23) to identify it as the first SSS to be driven by accretion from a helium star.  Such a system is potentially of great interest in the study of supernova type Ia progenitors, as one of the evolutionary routes is the SD (single-degenerate) channel in which a WD is accreting at very high rates \citep[see][and references therein]{Ruiter2020} from a highly evolved donor.  Furthermore, G23's analysis of archival data on \src\ from ROSAT through to XMM and eROSITA indicated that it's X-ray emission and spectrum appeared remarkably stable throughout.

From their inferred WD radius, G23 calculated its mass to be $M_1 = 1.20^{+0.18}_{-0.40} {\rm M_\odot}$ and that it was accreting material via an accretion disc fed by a helium star. Their long-term photometry revealed modulations at 1.1635 days, and its weak harmonic at 2.327 days, which were identified as possible orbital periods, and high resolution spectra appear to indicate a low inclination angle binary. 

The source of the supersoft X-ray component, however, has been challenged by \cite{Kato2023}, who point out that steady He burning on the WD surface requires an $\dot{M}_{\rm acc} \sim 10^{-6} {\rm ~M_\odot~yr^{-1}}$, which is much higher (by almost a factor 10) than that inferred by G23's observed \Lx . While G23 argue that stable burning at lower accretion rates is possible for rapidly rotating WDs \citep{Yoon2004, Wong2019}, \cite{Kato2023} offer a very different explanation.  They suggest that \src\ is a helium nova similar to V445 Pup \citep[e.g.][]{Kato2003, Woudt2005} that erupted $\sim 28$ years ago, and is currently being observed during the decay phase.

We therefore undertook an observational campaign to look for evidence of \src\ being a possible decaying helium nova, to confirm the absence of hydrogen in the system, and to constrain \Porb\ through extensive phase-resolved spectroscopy, as well as further photometry.

\section{Observations and data reductions}

Optical spectroscopic and photometric observations of [HP99] 159 were conducted between December 2022 and October 2024 at the Sutherland station of the South African Astronomical Observatory (SAAO). The photometric data was supplemented with OGLE-IV data.

\subsection{Spectroscopy}

\subsubsection{SALT/HRS}
High resolution spectroscopy was performed with the 11-m Southern African Large Telescope (SALT; \citealp{SALT2006}) equipped with a dual-beam fiber-fed echelle high resolution spectrograph (HRS; \citealp{Bramall2010, Bramall2012,  Crause2014}). The observations were conducted under program 2021-2-LSP-001 (PI: David Buckley) on 21 epochs between December 2022 and February 2024. The HRS 
covers the spectral range of 3740–8780 \AA ~(blue camera $\lambda$ = 3740–5560 \AA, red camera $\lambda$ = 5450–8780 \AA). It has three resolution modes, namely low (R = 14000–15000), medium (R=40000–43000) and high (R=67000–74000). Our observations were carried out in low and medium resolution modes, with Table~\ref{tab:hp99_spec_obs} providing more details.

\begin{table}
\begin{center}
\footnotesize
\caption{SALT/HRS spectroscopic observing log of \src\ } 
\label{tab:hp99_spec_obs}
\begin{tabular}{lccr@{~}c@{~}l}
\hline
Date & HJD (Start) & Resolution & $n_{\rm data}$ & $\times$ & $t_{\rm exp}$ \\ 
 & (2450000+) & Mode & & (s) & \\
\hline
 2022-12-26, 27, 30, 31 & 9940-9945  & Low & 1 & $\times$ & 2400 \cr
 2023-01-01 & 9946 & Low & 1 & $\times$ & 2400 \cr
 2023-01-12, 13, 14 & 9957-9959 & Low & 1 & $\times$ & 2400 \cr
 2023-01-30, 31 & 9975-9976  & Low & 1 & $\times$ & 2400 \cr
 2023-02-09 & 9985  & Low & 1 & $\times$ & 2400 \cr
 2023-02-12, 13 & 9988-9989  & Low & 1 & $\times$ & 2400 \cr
 2023-03-02 & 10006 & Low & 1 & $\times$ & 2400 \cr
 2023-03-12 & 10016 & Low & 1 & $\times$ & 2400 \cr
 2023-03-17 & 10021 & Low & 1 & $\times$ & 2400 \cr
 2023-11-17 & 10266 & Medium & 2 & $\times$ & 1800 \cr
 2024-01-31 & 10341 & Low & 1 & $\times$ & 2500 \cr
 2024-02-05, 08, 09 & 10346-10350  & Low & 1 & $\times$ & 2500 \cr
\hline
\multicolumn{6}{l}{\footnotesize{\textit{Note}: The number of data frames is $n_{\rm data}$ and the exposure time}}\\
\multicolumn{6}{l}{\footnotesize{~~~~~~~~~~per frame is $t_{\rm exp}$.}}\\
\end{tabular}
\end{center}
\end{table}

The primary HRS data reduction was performed with the SALT pipeline PySALT \citep{Crawford2010}, which included bias subtraction, cross-talk correction, plus gain and overscan correction. The MIDAS HRS automatic pipeline\footnote{https://astronomers.salt.ac.za/software/hrs-pipeline/} \citep{Kniazev2016, Kniazev2017} was used to perform the spectroscopic reductions which included flat-field corrections, order extraction, wavelength calibration with arc spectra of ThAr+Ar lamps, sky subtraction, and 2D to 1D spectral extraction. 

\subsubsection{1.9-m telescope and SpUpNIC}
\src\ was observed on 3 October 2024 with the SAAO 1.9-m telescope equipped with the SpUpNIC Cassegrain spectrograph ~\citep{Crause2016} using grating 4, giving a resolution of 1.8~\AA\ with the 1.5 arcsec slit width.  A CuAr lamp provided wavelength calibration.  Three science frames were obtained with an exposure time of 1200 sec for each. 

Standard {\small IRAF} routines \citep{Massey1997} were used to perform bias and flat field correction, and the spectra were wavelength-calibrated using the CuAr arc frames.  Flux calibration was performed with respect to the spectrophotometric standard star, LTT 7987, observed on the same night. The spectra were co-added to obtain one average spectrum.

\subsection{Photometry}
The SAAO 1.0-m telescope equipped with SHOC, the Sutherland High-speed Optical Camera\footnote{Further details on SHOC can be obtained from {https://www.saao.ac.za/science/facilities/instruments/shoc/}.} \citep{Coppejans2013} was utilized for photometry of \src. The observations were performed in conventional mode i.e. utilizing a 1\,MHz 16-bit conventional amplifier, with a pre-amp gain of 2.5 and 2x2 binning. No filters were used, meaning that \src\ was observed in ``white light''. The details of these observations are summarized in Table~\ref{phot_obs}.

\begin{table}
\begin{center}
\footnotesize
\caption{SAAO 1.0-m/SHOC photometric observing log of \src\ } 
\label{phot_obs}
\begin{tabular}{lcc}
\hline
Date & HJD & $t_{\rm exp} \times n_{\rm data}$\\ 
 & (2460000+) & (s) \\
\hline
2024-01-03  & 313 & 20$\times$426 \cr
2024-01-04  & 314  & 15$\times$987 \cr
2024-01-05  & 315  & 15$\times$818 \cr
2024-01-06  & 316 & 15$\times$726 \cr
2024-01-08  & 318  & 15$\times$200 \cr
2024-01-17  & 327  & 15$\times$726 \cr 
2024-01-18  & 328 & 15$\times$522 \cr 
2024-01-19  & 329 & 15$\times$726 \cr 
2024-01-20  & 330 & 15$\times$420 \cr  
2024-01-21  & 331  & 15$\times$576 \cr
2024-01-22  & 332  & 15$\times$242 \cr
2024-01-23  & 333  & 15$\times$484 \cr 
2024-01-24  & 334 & 15$\times$726 \cr 
2024-01-29  & 339 & 30$\times$81 \cr 
\hline
\multicolumn{3}{l}{\footnotesize{\textit{Note}: The exposure time per frame is $t_{\rm exp}$, and }}\\
\multicolumn{3}{l}{\footnotesize{~~~~~~~~~$n_{\rm data}$ is the number of data frames.}}\\
\end{tabular}
\end{center}
\end{table}

TEA-P{\small HOT}\footnote{https://bitbucket.org/DominicBowman/tea-phot/}, a Python-based pipeline developed by \cite{Bowman2019} specifically for SHOC data cubes, was exploited to perform our data reductions. Although capable of producing differential light curves from CCD reductions and adaptive elliptical aperture photometry, we only used it to slice the data cubes, obtaining the correct time stamps per frame (including the Barycentric Julian Date in the Dynamical Time standard at mid-exposure), and to perform bias subtraction and flat-field correction. 

\src\ is located in the LMC, a crowded stellar field, and so we used the {\small DAOPHOT} package in {\small IRAF} to perform PSF (Point-Spread Function) photometry. A weighted differential photometry procedure was applied, following the same method as discussed by \cite{Everett2001} and \cite{Burdanov2014}. Two comparison stars of similar brightness to the target, and which had the lowest mean variance were utilized.

\subsection{OGLE-IV photometric data}
The Optical Gravitational Lensing Experiment (OGLE) is a long-term, large-scale photometric survey focused on sky variability, including detecting microlensing events, and has been operational from the early 1990s \citep{Udalski1992}. It regularly monitors hundreds of millions of stars in the Galactic Bulge, Galactic Disc, and Magellanic Cloud regions and, hence, monitored [HP99] 159 during the fourth phase of the project (OGLE-IV). See \cite{Udalski2015} for a detailed description of the observing setup and data calibration. Here, we analyze OGLE-IV observations collected between 2010 and 2024. The data for [HP99] 159 were obtained in the $I$-band filter with typical exposure times of 150\,s. The photometry was performed using the difference image analysis technique \citep{Alard1998} as implemented by \citet{Wozniak2000}. The DIA pipeline implemented by OGLE returns underestimated photometric errors, so we applied the formula $\sigma_{\rm new} = \sqrt{(\gamma \times\sigma_{\rm old})^2 + \epsilon^2}$ (where $\gamma = 1.058$, $\epsilon = 0.0051$ for the OGLE field containing \src) specified by \citet{Mroz2024} to correct the OGLE-IV magnitude errors. Note that this represents a statistical error and does not fully capture the intrinsic uncertainty associated with the scatter.

\section{Results}

\subsection{Photometric variability}

\begin{figure}  
\centering
\includegraphics[width=0.47\textwidth]{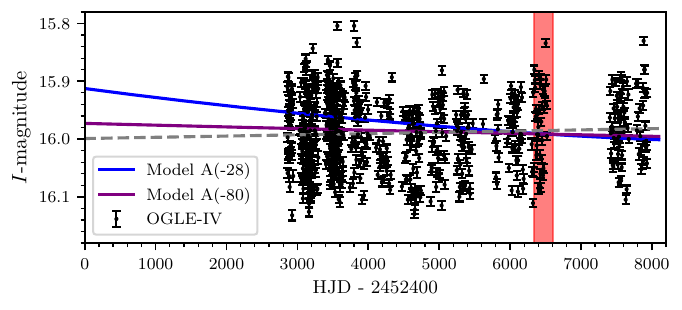}
\caption{OGLE-IV $I$-band light curve of \src\, obtained Mar 2010 -- Jan 2024, with a linear fit (dashed line) giving a mean of $I$=16, and demonstrating no detectable decline over the last 14 yrs.  The blue and purple lines represent the 28 yr and 80 yr decaying nova models, respectively, of \citet{Kato2023}. The red region indicates the time interval of the XMM-Newton and eROSITA observations.}
\label{fig:ogle}
\end{figure}

The 14-year OGLE-IV $I$-band light curve (Figure~\ref{fig:ogle}) shows that \src\ remained within $\pm$0.1 mag of $I\sim16$ throughout this interval. A linear model fitted to the data indicated a slope of $-0.0007 \pm 0.0006 {\rm ~mag~yr^{-1}}$ (brightness increase) and a horizontal shift of $15.9897 \pm 0.0027$ mag. A t-test indicated that the null hypothesis slope of zero at a 3$\sigma$ confidence level (99.73\%) cannot be rejected. This provides evidence that the system has not declined over time. 

We also compared the OGLE-IV light curve with the decaying nova model (and a slow X-ray decay of $\tau_{SSS} = 120$ yr) proposed by \cite{Kato2023}.  The X-ray light curve of the decaying nova model was converted to $M_{\rm V}$ using the correlation $M_{\rm V} = 0.83 - 3.46 \log\Sigma$, where $\Sigma = (L_{\rm X}/L_{\rm Edd})^{1/2}(P/1{\rm hr})^{2/3}$, derived by \citet{vanTeeseling1997} for SSS.  Since the X-ray light curve only accounts for the SSS contribution, a He star flux was added so that the total system brightness scaled to fit through the OGLE-IV data at the time of the XMM-Newton and eRosita observations.  The total contribution $M_{V}$ was converted to $V$-mag and a $V-I = 0.06$ offset (obtained from G23 Extended Data Fig.~3) was added.  The model is plotted in Figure~\ref{fig:ogle} using two age options for the nova outburst: 28 yr ago (blue line) and 80 yr ago (purple line).  

In the OGLE interval, the 28 yr decay model has an almost linear slope of $+0.00351 \pm 0.00001 {\rm ~mag~yr^{-1}}$ (brightness decrease) and a t-test rejects a null hypothesis slope of zero at a 3$\sigma$ confidence level (99.73\%).  The 80 yr decay model has a much shallower linear slope of $+0.000974 \pm 0.000001 {\rm ~mag~yr^{-1}}$ (brightness decrease) and a t-test indicates a null hypothesis slope of zero (3$\sigma$ confidence level; 99.73\%) cannot be rejected.  While the 80 yr decay model is consistent with the OGLE data, we reject it since the model fits very poorly to the observed SSS fluxes, as also noted by \citet{Kato2023}.  These results, together with the fact that the blackbody temperature has also been stable for the past 30 years, does not support the decaying helium nova scenario, as even the most shallow decline consistent with the observations in their Fig.~3 would have been detectable in the OGLE data.

\begin{figure} 
\centering
\includegraphics[width=0.47\textwidth]{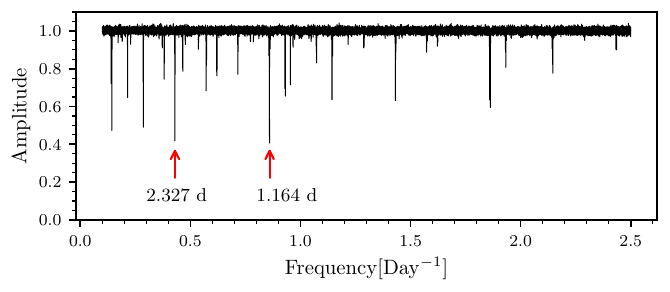}
\caption{PDM periodogram of the OGLE-IV light curve (Fig.~\ref{fig:ogle}) in the frequency range of $0.1 < f < 2.5 {\rm ~cycles~day^{-1}}$.  The two strongest signals are marked, and are the same as those seen in our and G23's Lomb-Scargle periodograms.}
\label{fig:ogle_pdm}
\end{figure}

As in G23, we performed a Lomb-Scargle (LS) analysis of the OGLE-IV light curve, and also strongly detected modulations at $P = 1.1635 \pm 0.0001$ d and a weaker signal at $P = 2.3269 \pm 0.0003$ d (our periodogram is essentially identical to G23's Fig.~4a). Since G23 favoured the 2.3\,d signal as a possible orbital period (due to its lower variance), we decided to perform a phase dispersion minimization (PDM) period analysis, as it is insensitive to the light curve shape, and is able to determine periods with high accuracy.  We used the \textit{pdm} task in the Starlink PERIOD package (Version 5.0-2), selecting 20 bins (each of width 0.11\,d) and a frequency range of $0.1 < f < 2.5 {\rm ~cycles~day^{-1}}$. The resulting PDM is shown in Figure~\ref{fig:ogle_pdm} and clearly detected both modulations, with them now being of essentially equal strength. 

\begin{figure} 
\centering
\includegraphics[width=0.47\textwidth]{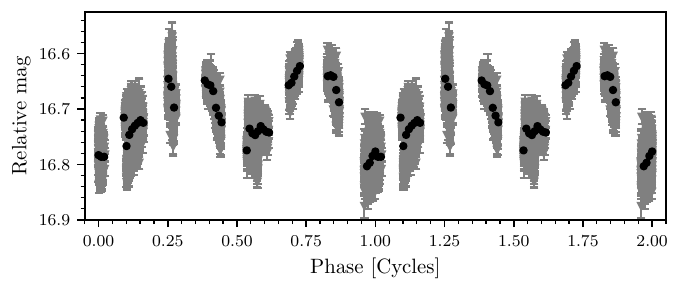}
\caption{The phase-folded light curve on the proposed 2.33 d orbital period obtained with the SAAO 1.0-m/SHOC (3 -- 29 Jan 2024). Two cycles are shown for clarity.}
\label{fig:saao_shoc_folded}
\end{figure}

The light curve of \src\ obtained with the SAAO 1.0-m/SHOC (3 -- 29 Jan 2024) was phase-folded on the suggested \Porb\ of 2.327 d (Figure~\ref{fig:saao_shoc_folded}), using BJD2455255.443 as $T_0$. It displays the same double-humped light curves typically seen in accreting close binary systems, and is in agreement with that of G23.  We therefore also support a \Porb\ of 2.327\,d.

\subsection{Spectroscopic variability}

\begin{figure} 
\centering
\includegraphics[width=0.47\textwidth]{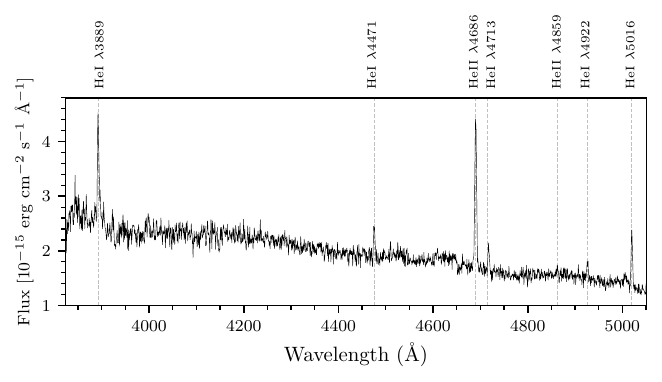}
\caption{SAAO 1.9-m/SpUpNIC spectrum of \src\ on 3 Oct 2024. All the emission lines are attributed to \he{i} and \he{ii}, confirming the absence of hydrogen.}
\label{fig:HP99_spupnic}
\end{figure}

The 1.9-m/SpUpNIC spectrum (Figure~\ref{fig:HP99_spupnic}), covering the wavelength range $\lambda \sim$ 3820 -- 5050~\AA, is very similar to that of G23 Fig.~2. In particular, this purely emission-line spectrum shows no hydrogen (Balmer) lines, only those attributable to \he{i} and \he{ii}.

\begin{figure} 
\centering
\includegraphics[width=0.48\textwidth]{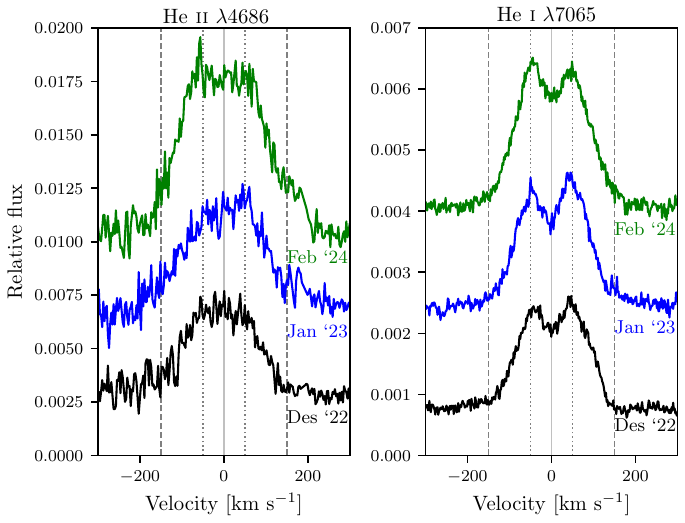}
\caption{\he{ii} and \he{i} line profiles in SALT/HRS spectra, obtained between Dec 2022 and Feb 2024. The plots are offset vertically by an arbitrary constant.}
\label{fig:He_line_profiles}
\end{figure}

The high resolution spectra obtained with SALT/HRS also support the results of G23, as seen in the detailed \he{ii} and \he{i} line profiles of Figure~\ref{fig:He_line_profiles}. They are all double-peaked He emission lines, evidence of an established accretion disc, together with changes in the asymmetry of the profiles that is typical, and often attributed to the hot spot (created by the mass-transfer stream impacting the disc) rotating with the orbital period. A peak separation of $\sim$100~\kms\ implies an outer disc velocity of $\sim$50~\kms\, and the wings of the emission lines indicate an inner disc velocity of $\sim$150~\kms. These values are low for accretion discs in close interacting binaries, indicating that the system is likely being viewed from a very low inclination angle if the lines arise only from the disk modulated by the Keplerian and orbital motions.\\

\subsection{Radial velocity curves}

\begin{figure} 
\centering
\includegraphics[width=0.47\textwidth]{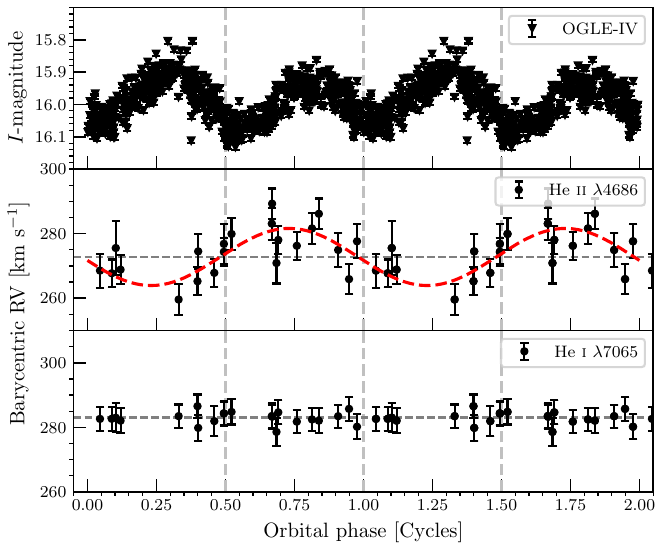}
\caption{\textit{Top panel:} The phase folded OGLE-IV light curve. \textit{Middle panel:} \he{ii} $\lambda 4686$ radial velocity curve. A sine fit (red dashed line) gives $K_1 \sim 9 \pm 1 {\rm ~km~s^{-1}}$.  \textit{Bottom panel:} \he{i} $\lambda 7065$ radial velocity curve. Two cycles are shown for clarity. Error bars show $\pm 1 \sigma$.}
\label{fig:folded_rv}
\end{figure}

With our extensive SALT spectroscopy of \src\ (Table~\ref{tab:hp99_spec_obs}) we could study the radial velocity (RV) curves of these emission lines. To measure $K$, we applied the method of \cite{Shafter1983}. The wings of the two strongest emission lines, \he{ii} $\lambda$4686 and \he{i} $\lambda$7065, were fitted with Gaussian functions to determine their central wavelengths. The resulting barycentric-corrected RV curves of \he{ii} and \he{i} are plotted in the middle and bottom panels of Figure~\ref{fig:folded_rv}, respectively, along with the OGLE-IV light curve (top panel), all phase-folded on \Porb\ = 2.327\,d. A sine function was fitted to the \he{ii} RV curve (red dashed line), giving a systemic velocity of $\gamma = 273 \pm 1$ \kms\ and a semi-amplitude of the accretion disc of
$K_{disc} = 9 \pm 1$ \kms. 
The \he{i} RV data points clearly show no significant modulation, but interestingly do show a mean RV offset of $\sim$+10 \kms\ with respect to \he{ii}. The lack of RV modulation for \he{i} indicates that the \he{ii} emission is originating from a region close to the WD, while the \he{i} likely comes from further out in the accretion disc, where the velocities are lower. 

Assuming a circular orbit and the white dwarf located at the centre of an axisymmetric accretion disc, then the $K_1$ value for the WD (assumed = $K_{disc}$) together with \Porb\ = 2.327\,d gives us the donor's mass function 
\begin{equation}
\frac{{M_2}^3}{\left(M_1 + M_2\right)^2}{\rm ~sin}^3 i = \frac{P_{\rm orb}}{2\pi G}~K_1^3 
\end{equation}

i.e. $f(M_2) = 1.75{\times}10^{-4} {\rm ~M_\odot}$  and is clearly indicative of a low-inclination system ($i<5^\circ$ for $M_2{\geq}1 {\rm ~M_\odot}$).
Furthermore, from \Porb\ we know from Kepler's 3rd Law that the size of the binary must be $a = 7.4 (M_1 + M_2)^{1/3}~{\rm R_\odot}$.

\section{Discussion}

\subsection{Binary parameters}

While the WD mass was constrained by G23 to be $M_1 = 1.20^{+0.18}_{-0.40}{\rm ~M_\odot}$ based on the measured X-ray blackbody radius, the absence of hydrogen in the disc led them to assume that it must be accreting from a He star, which must be filling its Roche-lobe in order to drive the necessarily large mass transfer rates.  This implies a He star donor with mass in the range $M_2 \sim 0.8 - 2.0{\rm ~M_\odot}$ (see also \citealp{Kato2023} who derive similar masses). For unstable mass transfer proceeding on a thermal timescale, thereby driving a high accretion rate \citep{Kahabka2006} means that the mass ratio $q (= M_2/M_1) \gtrsim$1 for the inferred  massive WD.  However, to reproduce the low RV semi-amplitude, such masses also imply a very low orbital inclination, namely $i \sim 3.5 - 4.8$ degrees. \\

Given that the He star must be filling its Roche-lobe, we can use the \cite{Eggleton1983} approximation to:
\begin{equation}
\frac {R_2}{a} = \frac{0.49q^{2/3}}{0.6q^{2/3} + {\rm ln}\left(1+q^{1/3}\right)}~~~ {\rm for}~~~ 0 < q < \infty,
\end{equation}
to estimate the He star radius (and by replacing $q$ with $q^{-1}$, we can also calculate the WD's Roche lobe radius, $R_1$). This yields the binary parameters given in Table~\ref{tab:binary}.

\begin{table}
    \centering
    \footnotesize
     \caption{Estimated binary parameters of \src\ for three donor masses}
    \label{tab:binary}   
    \begin{tabular}{l|c|c|c}
       Parameter & $M_2 = 0.8 {\rm ~M_\odot}$ & $M_2 \geq 1.2 {\rm ~M_\odot}$ & $M_2 = 2 {\rm ~M_\odot}$ \\
       \hline
        $M_{\rm 1}$ (${\rm M}_{\odot}$) & 1.2 & 1.2 & 1.2  \\
        $q~(=M_{\rm 2}/M_{\rm 1}$) & 0.67& $\geq$ 1 & 1.7 \\ 
        $R_{\rm 1}$ (${\rm R}_\odot$) & 3.9 & $<$ 3.8 & 3.7  \\
        $R_{\rm 2}$ (${\rm R}_\odot$) & 3.2 & $>$ 3.8 & 4.6  \\
        $i$ (degrees) & 6.4 & $<$ 4.8 & 3.5\\        
        \hline
    \end{tabular}
\end{table}

Helium stars with these properties have been modelled by \citet{Gotberg2018}, with their Fig.~2 (see also \citealp{Kato2023}, Fig.~4) indicating that a $\sim 4{\rm ~M_\odot}$ star will enter the mass transfer phase when it has expanded to sizes comparable to that calculated in Table~\ref{tab:binary}.  We also chose such relatively low mass He stars based on the optical model spectra presented in their Fig.~5.  More massive stars ($>7{\rm ~M_\odot}$) can be excluded, as they would exhibit much more powerful, broader \he{ii} emission lines that are typical of Wolf-Rayet stars.  The much narrower, weaker emission features seen in our spectra are almost certainly driven by the SSS component, and not the donor.  Indeed, the model He star spectra of \citet{Gotberg2018} in the lower mass range show only weak Balmer absorption features, and even they would be undetectable as a result of the mass-transfer stripping of the hydrogen envelope, leaving the bare He star to build up a helium-dominated accretion disc that feeds the WD.

\subsection{Can we believe the low inclination?}

There are a number of reasons why we should treat such a low inferred {\it i} value from the radial velocity curve with caution, although there are important caveats.

\begin{enumerate}
    \item on statistical grounds, pole-on binaries are rare.  It is well-known (based on solid angles on the sky) that a random distribution of orbital axis orientations has a chance of {\it i} values that is proportional to ${\sin}i$;
    
    \item however, there may also be observational selection effects at work.  One of these is the ease with which the very low energy SSS X-rays can be absorbed by intervening matter, and this is particularly true at higher binary inclinations where the mass transfer stream and disc material itself can block the SSS component.  This means they are more readily detectable when observed at lower inclinations;
    
    \item such a low {\it i} is not consistent with our observed photometric modulation  (Figure~\ref{fig:folded_rv} top panel).  \src\ could be considered as a contact binary, and the simulated light curves of \citet{Rucinski01} imply that $i\sim$40$^{\circ}$ is needed to produce an $\sim$0.1 mag amplitude modulation;

    \item the prototypical SSS RX\,J0513.9-6951 has a similar \he{ii} emission line RV curve \citep{Southwell96} with a low $K_1$ of 14.5\,km\,s$^{-1}$ that would infer $i <$7$^{\circ}$, yet it also displays weaker satellite emission lines displaced by $\pm$3900\,km\,s$^{-1}$ from the main feature. These are interpreted as jets emitted perpendicular to the plane of the disc, travelling with at least the WD escape velocity, again arguing against a low inclination, as also indicated by its photometric light curve;

    \item our double-peaked \he{i} line profile in Figure~\ref{fig:He_line_profiles} is a classic feature of accretion discs in CVs and LMXBs, and is taken as arising from advancing and receding material towards the outer disc.  Again, this would not be expected in a very low-inclination system.
\end{enumerate}

\subsection{Light curve modelling}

Since the simulated light curves of \citet{Rucinski01} suggest that we should be considering higher inclination values, we decided to investigate this in more detail by using {\small PHOEBE} (Version 2.4.21; \citealp{Prsa2016}) to model the OGLE-IV light curve of \src. {\small PHOEBE} is a software package that is used for eclipsing binaries, but does not include accretion physics.  However, given that the ``steady'' SSS requires ongoing high levels of mass transfer into the accretion disc around the WD (as depicted in \citealp{Hachisu03}, Fig.3) we have approximated this as an A0 star orbiting the He star, and used {\small PHOEBE} to build synthetic light curves at different $i$.  We note a similar approach was used in \citet{Smak2001} in modelling the double-humped light curve of V Sge, a high mass transfer cataclysmic variable.  The accretion discs in such systems are bloated and cannot be modelled as standard accretion discs (see e.g. \citealp{Hachisu03}).

The simulated system was treated as a contact binary with the stars in circular orbit. The binary parameters \Porb\ = 2.327\,d, $a= 10.9~\rm{R}_\odot$ and $q= 1.67$ were fixed and we assumed an A0 star ($M_1 = 1.2~{\rm M_\odot}$, $T_{\rm eff} = 10000$ K) to represent the WD+hot disc. An effective temperature of $T_{\rm eff} = 15850$ K was set for a $M_2 = 2~{\rm M_\odot}$ He star. The stellar radii were set to the respective Roche lobe radii ($R_1 = 3.66~\rm{R}_\odot$ and $R_2 = 4.6~\rm{R}_\odot$). 

\begin{figure} 
\centering
\includegraphics[width=0.48\textwidth]{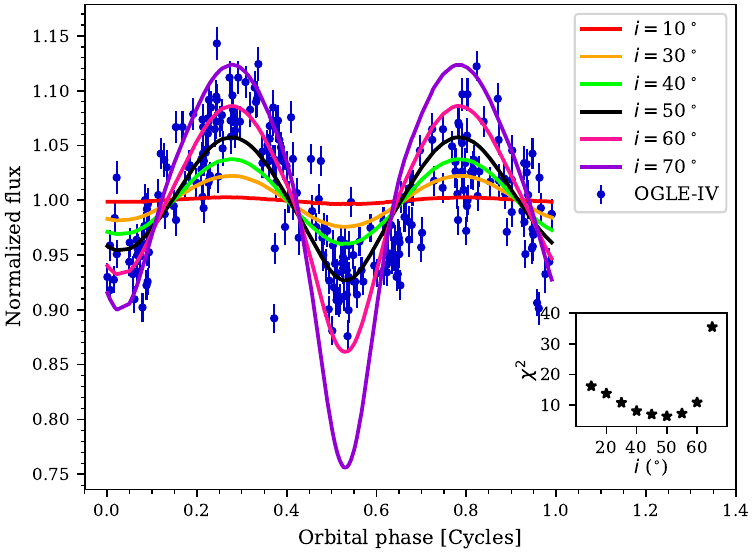}
\caption{The phase folded OGLE-IV light curve with the synthetic light curves produced with {\small PHOEBE} at different inclination angles. \textit{Insert}: Reduced $\chi^2$ values for the OGLE-IV light curve fitted with the synthetic light curves at different inclination angles.}
\label{fig:synthetic_lcs}
\end{figure}

The OGLE-IV light curve was converted to flux (using the Vega flux zeropoint\footnote{https://github.com/sczesla/PyAstronomy}; \citealp{Czesla2019}) and normalized, and plotted with the synthetic light curves at different $i$ in Figure~\ref{fig:synthetic_lcs}. A $\chi^2$ analysis indicates that the synthetic light curve with $i = 50^\circ$ fits the OGLE-IV data the best with a reduced $\chi^2 = 6.4$ (Figure~\ref{fig:synthetic_lcs} insert). As {\small PHOEBE} does not model accretion discs and hot spots, it cannot model the excess emission above the synthetic light curve at phase 0.25 contributed by the hot spot in \src.  

This does leave us with conflicting indications for $i$, but based on the unlikelihood of pole-on systems we believe the higher $i$ implied by the light curves to be more likely.  We then assume that the \he{ii} must be produced in outflowing material further above, or lost from, the disc and hence will display a diluted RV modulation. This is supported by the $\gamma$ velocity of the \he{ii} RV curve being blue-shifted with respect to \he{i}.

\subsection{[HP99]~159 as an SSS}

It has already been noted that the observed, and steady, \Lx\ can be explained by a mass accretion rate of \Mdot\ $\sim 1.5 \times 10^{-7} {\rm ~M_\odot~yr^{-1}}$, but the SSS paradigm requires steady He-burning on the WD surface and that needs a mass-transfer rate significantly ($\sim$5-10$\times$) higher.  So, is \src\ really comparable to other SSS in the Magellanic Clouds and the Milky Way?  Firstly, we compare their observed properties in Table~\ref{tab:SSSprops}, where it is clear that while \src\ is at the faint end of the list, it is actually slightly brighter than CAL87 (as expected given its lower $i$).  So this issue is not unique to \src.

This comparison can be taken further by recognising that in the canonical SSS the dominant light source is the accretion disc, and that will be heated by the close-to-Eddington limited SSS emission from the WD.  This makes them analogous to their LMXB (low-mass X-ray binary) cousins, where \citet{vanParadijs1994} (hereafter vPM94) had shown that there was a correlation between their $M_{\rm V}$ and a combination of their \Lx\ and \Porb\ (the latter of course links to the size of the accretion disc).  That correlation was $M_{\rm V} = 1.57 - 2.27 \log\Sigma$, where $\Sigma = (L_{\rm X}/L_{\rm Edd})^{1/2}(P/1{\rm hr})^{2/3}$, and it was used by \citet{vanTeeseling1997} to show that a very similar curve fits the SSS as well.

\begin{figure} 
\centering
\includegraphics[width=0.48\textwidth]{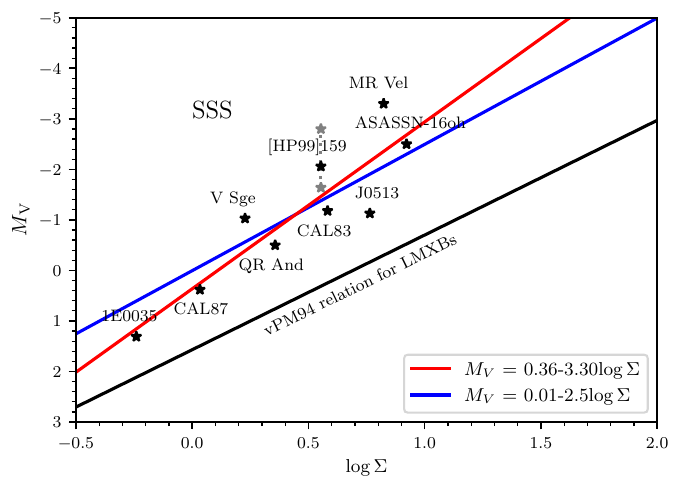}
\caption{M$_V$ against $\log\Sigma$ for the luminous SSS in the Magellanic Clouds and the Milky Way (adapted from \citealp{vanTeeseling1997}, Fig.~3). The red line is a straight-line fit for the SSS and the blue line is a fit with the slope fixed at -2.5. The black line is not a fit, but simply the relation for LMXBs ($M_{\rm V} = 1.57 - 2.27 \log\Sigma$) derived by vPM94 assuming the emission is driven by X-ray irradiation of the disc.} 
\label{fig:vTplot}
\end{figure}

We take the opportunity here to update this relationship by including \src\ and exploiting the now more accurate galactic distances available from Gaia DR3.  The new $M_{\rm V} - \Sigma$ plot for the SSS in Table~\ref{tab:SSSprops} is shown in Figure~\ref{fig:vTplot}, together with a simple straight-line fit (the red line), as well as the original vPM94 black line that fits the steady LMXBs.  As they noted (and also \citealp{Revnivtsev12}), since $\Sigma$ is a flux, then the slope of this line should be --2.5, and accordingly we show (blue line) the fit with the slope so fixed.  This demonstrates an important point.  The SSS as a group are clearly displaced above the LMXBs, and so have, for a given value of $\Sigma$, an $M_{\rm V}$ that is 1.5 mags brighter.  We believe that this is due to the much larger discs and donors of SSS relative to the LMXBs, essentially doubling the irradiated areas, as is clearly demonstrated by the schematic diagram (Fig. 3) in \citet{Hachisu03}.

However, \src\ is different from all other SSS in that it has a He star donor, which should be optically luminous in its own right, contributing significantly to the $M_{\rm V}$ of --2.8 for \src.  From \citet{Gotberg2018} and \citet{Kato2023} we estimate that a 0.8~$\rm M_\odot$ He star will have $M_{\rm V}$ = --2.03, and this leads to $M_{\rm V}$=--2.06 for the SSS component alone, which is the value we have used in Figure~\ref{fig:vTplot} (see also \citealp{Kato2023}).  The gray star above \src\ indicates the SSS component $M_{\rm V}$ = --1.83 if the He star has a $M_{\rm V}$ = --2.23 (determined from the 28 yr decay model) and the gray star below \src\ indicates the SSS component $M_{\rm V}$ = --1.64 if the He star has a $M_{\rm V}$ = --2.34 (determined from the 80 yr decay model).  This makes it quite consistent with the other known SSS.  We also note that the \citet{Hachisu03} interpretation indicates likely considerable mass outflows from the system. Outflows out of the plane of the orbit will show a much lower radial velocity modulation against orbital phase, which could explain our spectroscopic observations.

\begin{table}
	\centering
	\footnotesize
		\caption{SSS Key Properties}
		\label{tab:SSSprops}
		\begin{tabular}{l@{~~~}c@{~~~}c@{~~~}c@{~~~~~}l} 
			{Source} & {$V$} & {log$L_{\rm X}$} & \Porb & {Notes}\\
		& (range)& (\ergs\ ) & (d) & \\
					\hline
                 {Magellanic Clouds} & & & & \\
                                        \hline
			CAL83 & 16.2--17.1 & 37.3 & 1.04 & 450 d on-off cycle\\
			CAL87 & 19--21 & 36.7 & 0.44 & eclipsing\\
			RX\,J0513.9-6951 & 17.4 & 37.8 & 0.76 & 170 d on-off cycle\\
			1E\,0035.4-7230 & 20.3 & 37.3 & 0.17 & short \Porb\\
			ASASSN-16oh & 16.5--20.5 & 37.0 & 5.6 & $\sim$1\,yr outburst\\
                                        \hline
                        \src\ & 16 & 36.8 & 2.33 & this paper\\
                                        \hline
                 {Galactic} & & & \\
                                        \hline
                        QR\,And & 11.5--12.6 & 37.1 & 0.66 & E(B-V)=0.10\\                
                        V\,Sge & 10.5--14 & 36.7 & 0.51 & E(B-V)=0.33\\
                        MR\,Vel & 17.9 & 38.0 & 4.03 & E(B-V)=2.1\\                 
                                        \hline
                 
		\end{tabular}
\end{table}

\subsection{Comparison with other He star binaries}

We investigated whether \src\ is comparable to other known systems. In fact, there are a few WD-He star binaries that are of interest as SN Ia candidates, namely KPD\,1930+2752 \citep{Geier2007}, V445\,Pup \citep{Kato2003}, HD\,49798 \citep{Brooks2017}, CD-$30^{\circ}$11223 \citep{Vennes2012} and PTF1\,J2238+7430 \citep{Kupfer2022}. Of particular interest from this group is V445\,Pup, which underwent a nova ouburst in 2000, and remarkably also showed no hydrogen lines in its spectrum, making it the first He nova \citep{Kato2003}, and hence potentially analogous to \src. Interestingly, \cite{Schaefer2025} shows that the WD of V445\,Pup appears to be losing mass after each eruption, making it less likely to undergo a SNIa explosion.

However, there were two forms of He star companion considered (see e.g. \citealp{Woudt2009} and references therein) for V445\,Pup, one a non-degenerate, evolved He star, the other a fully degenerate He WD (also known as AM\,CVn systems).  These latter, double-degenerate binaries, have very short orbital periods (usually $\leq$60 mins).  This is significant, as very recently, \citet{Schaefer2025} has proposed an orbital period for V445\,Pup of 1.87\,d, based on long-term archival (pre-nova) optical photometry.  But his resulting folded light curve is very similar in shape and amplitude to that of \src\ when folded on half the orbital period, and so we suspect that the true period of V445\,Pup is actually 3.75\,d.  More importantly, such a long period automatically rules out the AM\,CVn scenario, and so V445\,Pup is definitely worth comparing more closely with \src.  

\subsection{Binary evolution}
From the properties of the expanding nebula surrounding V445\,Pup, \cite{Woudt2009} have derived a precise distance of 8.2$\pm$0.5\,kpc, and estimate an E(B-V) of 0.62 (and hence an A$_V{\sim}$2.5).  Furthermore, there are strong arguments for a massive WD ($\sim1.35{\rm ~M_\odot}$) and \citet{Kato2008} indicate that $M_2>0.8{\rm ~M_\odot}$, which, combined with \Porb\ = 3.75\,d, gives $R_2=4.7{\rm ~R_\odot}$.  These are very similar parameters to those we have derived for \src\ in Table~\ref{tab:binary}, and are consistent with what is expected for He star evolution as described in \cite{Gotberg2018} and \cite{Kato2023}. 

Regarding the actual binary system evolution, we note that \cite{Podsi10} has modelled binary evolution in the ``supersoft'' channel of potential SNIa systems.  In particular, his Fig.\,2 produces, after $\sim$2$\times$10$^6$\,yrs, a $\sim 1 {\rm ~M_\odot}$ WD accreting at $\sim$10$^{6-7}{\rm ~\dot{M}_\odot~yr^{-1}}$ from a donor of $M_2, R_2 = 2.1{\rm ~M_\odot}$, $4.6{\rm ~R_\odot}$.  This was based on the observed properties of the recurrent nova U\,Sco, which also bear comparison with \src, especially since it was found \citep{Johnston92} to be ``H-deficient''.  Also targeting U\,Sco, \cite{Hachisu99} produce a system with a $\sim 1{\rm ~M_\odot}$ WD accreting from a $\sim2-3.6{\rm ~M_\odot}$ companion with a $\sim 0.5-5$\,d orbital period (see their Fig.\,1.F).  Of relevance to \src\ is that the donor has a ``He-rich envelope'' acquired from the pre-WD in an earlier phase of its evolution.

\section{Conclusions}

Our extensive spectroscopic and photometric study of \src\ has shown that:
\begin{itemize}
\item it has remained within $\pm$0.1\,mag of its mean $I \sim 16$ level over the last 14 years, with no indication of decay in the light curve;
\item together with the  constant blackbody temperature determined by G23, this provides strong evidence that \src\ is not in the decaying phase of a helium nova event as was suggested by \citet{Kato2023}, and that it should be considered as a comparable member of the group of Magellanic Cloud SSS;
\item a PDM period analysis of the OGLE-IV light curve and the phase-resolved spectroscopic results confirm that \Porb = 2.327\,d;
\item we see only helium emission lines in the spectrum, indicating that the donor was likely an (originally) $\sim 4{\rm ~M_\odot}$ He star; 
\item our \he{ii} RV curve yields $\gamma = 273 \pm 1$ \kms\ and $K_1 = 9 \pm 1$ \kms\, in conflict with our double-peaked orbital light curve regarding the binary $i$, which we believe to be closer to $i\sim50^{\circ}$;
\item we propose that substantial mass outflow from the system accounts for the low RV modulation amplitude;
\item we propose a donor mass range of $M_2 \sim 1.2 - 2.0{\rm ~M_\odot}$, very similar to the He nova V445\,Pup;
\item this confirms that \src\ is indeed the first SSS with an evolved helium donor and therefore a single-degenerate scenario SNIa progenitor;
\item we have demonstrated a clear $\sim$1.5\,mag difference in $M_{\rm V}$ of SSS systems when compared to the $M_{\rm V} - \Sigma$ properties of LMXBs.
\end{itemize}

\section*{Acknowledgements}

Firstly, we thank the anonymous referee for a constructive report that has improved the presentation of our results.  Some of the observations reported in this paper were obtained with the Southern African Large Telescope (SALT), under programme 2021-2-LSP-001 (PI: DAHB), the 1.0-m telescope and the 1.9-m telescope located at the Sutherland station of the South African Astronomical Observatory (SAAO).  Polish participation in SALT is funded by grant No. MNiSW DIR/WK/2016/07. This paper also used observations made by the Optical Gravitational Lensing Experiment (OGLE).  PAC would like to thank Patrick Woudt for useful discussions on V445\,Pup, and also acknowledges the Leverhulme Trust for an Emeritus Fellowship. DAHB acknowledges research support from the National Research Foundation. HS acknowledges research support from the Department of Higher Education and Training, RSA. The OGLE project has received funding from the Polish National Science Centre grant OPUS-28 2024/55/B/ST9/00447 to AU.

\section*{Data Availability}

The data underlying this article will be shared on reasonable request to the corresponding author.


\bibliographystyle{mnras}
\bibliography{references}







\bsp	
\label{lastpage}
\end{document}